\documentclass[12pt]{article}
\usepackage{amsmath}
\usepackage{amssymb}
\usepackage[dvips]{graphicx}
\usepackage{amsmath,amscd}
\usepackage{color}
\usepackage{mathrsfs}
\usepackage{latexsym}
\usepackage{bm}
\makeatletter

  \@addtoreset{equation}{section}
\makeatother
\allowdisplaybreaks[4]

\begin{document}
\begin{titlepage}
\thispagestyle{empty}
\begin{flushright}

\end{flushright}
\vspace{3mm}

\begin{center}
{\LARGE Dark energy from gravitational corrections}

\end{center}

\begin{center}
\lineskip .45em
\vskip1.5cm
{\large Yugo Abe$^a$\footnote{E-mail: yugo@riko.shimane-u.ac.jp}, 
Masaatsu Horikoshi$^b$\footnote{E-mail: 14st307b@shinshu-u.ac.jp}
and Yoshiharu Kawamura$^b$\footnote{E-mail: haru@azusa.shinshu-u.ac.jp}

\vskip 1.5em
${}^a\,${\large\itshape Graduate School of Science and Engineering, 
Shimane University, }\\
{Matsue 690-8504, Japan}\\[1mm]
${}^b\,${\large\itshape Department of Physics, Shinshu University, }\\
{Matsumoto 390-8621, Japan}
}

\vskip 4.5em
\end{center}
\begin{abstract}

We study physics concerning the cosmological constant problem
in the framework of effective field theory 
and suggest that a dominant part of dark energy can originate from
gravitational corrections of vacuum energy,
under the assumption that the classical gravitational fields
do not couple to a large portion of the vacuum energy effectively,
in spite of the coupling between graviton and matters
at a microscopic level.
Our speculation is excellent with terascale supersymmetry.

\end{abstract}
\end{titlepage}

\newpage

\abovedisplayskip=1.0em
\belowdisplayskip=1.0em
\abovedisplayshortskip=0.5em
\belowdisplayshortskip=0.5em
\section{\it Introduction}

The cosmological constant problem (CCP) is a biggest puzzle
in particle physics~\cite{CCP1,CCP2,CCP3,CCP4,CCP5,CCP6,CCP7},
and consists of several pieces.

The first piece is that the vacuum energy density $\rho_{\rm v}$ 
can be the cosmological constant $\varLambda_{\rm c}$
and various sources of $\rho_{\rm v}$ exist, 
e.g., a zero point energy of each particle and
potential energies accompanied with phase transitions
such as the breakdown of electroweak symmetry
via Higgs mechanism
and the chiral symmetry breaking due to quark condensations.

The second one is that $\rho_{\rm v}$ can receive
large radiative corrections including a cutoff scale.

The third one is that the experimental
value of $\varLambda_{\rm c}$ is estimated
as $\varLambda_{\rm c(exp)} \doteqdot 8 \pi G \cdot 
2.4 \times 10^{-47}$GeV$^4$
where $G$ is the Newton constant,\footnote{
In this article, we use the natural unit that
$c=1$ and $\hbar = 1$.
}
from the observation that the expansion of present universe 
is accelerating~\cite{DE}.
The energy density defined by 
$\rho_{\rm DE} \equiv \varLambda_{\rm c(exp)}/(8 \pi G)$
is referred as $\lq\lq$dark energy density'',
and its existence has been a big mystery.

The pieces of puzzle are not fitted
in the framework of the Einstein gravity
and the standard model of particle physics.
Because a fundamental theory including gravity
has not yet been established,
it would be meaningful to give a suggestion
based on an effective description
of various experimental results concerning
the vacuum energy.
It might be necessary to
introduce a principle, assumptions and/or a framework
beyond common sense of an accepted physics,
and then the CCP is replaced by the problem
to disclose the essence of new staffs.

In this article, 
we study physics on the CCP
in the framework of effective field theory 
and suggest that a dominant part of dark energy can originate from
gravitational corrections of vacuum energy
under the assumption that the classical gravitational fields
do not couple to a large portion of the vacuum energy effectively,
in spite of the coupling between graviton and matters
at a microscopic level.

The content of our article is as follows.
In the next section, we explain the pieces of puzzle on the CCP
and their implications.
In Sect. 3, we give an effective description 
for physics concerning the CCP and
predict that gravitational corrections of vacuum energy
can be a candidate of dark energy.
In the last section, we give conclusions and discussions.

\section{\it Pieces of puzzle}

\subsection{\it Vacuum energy density}

{\it The vacuum energy density $\rho_{\rm v}$ 
can be the cosmological constant $\varLambda_{\rm c}$.}
The energy-momentum tensor of perfect fluids is
given by
\begin{eqnarray}
T^{\mu\nu} = (\rho + p) u^{\mu} u^{\nu} + p g^{\mu\nu},
\label{T-pf}
\end{eqnarray}
where $\rho$, $p$ and $u^{\mu}$ are an energy density,
a pressure and a four-velocity of fluids, respectively.
The energy-momentum tensor of vacuum is of the form,
\begin{eqnarray}
\langle T^{\mu\nu} \rangle 
= - \frac{\varLambda_{\rm v}}{8 \pi G} g^{\mu\nu},
\label{T-v}
\end{eqnarray}
where $\varLambda_{\rm v}$ is a constant.

From (\ref{T-pf}) and (\ref{T-v}),
we obtain the relations,
\begin{eqnarray}
\rho_{\rm v} = - p_{\rm v},~~
\langle T^{\mu\nu} \rangle = - \rho_{\rm v} g^{\mu\nu},
\label{Rel-v}
\end{eqnarray}
where $\rho_{\rm v}(=\varLambda_{\rm v}/(8 \pi G))$ 
and $p_{\rm v}$ are an energy density and
a pressure of vacuum, respectively.
The vacuum has a negative pressure with $\rho_{\rm v} (>0)$.

The Einstein equation is given by
\begin{eqnarray}
R^{\mu\nu} - \frac{1}{2} g^{\mu\nu} R + \varLambda_{\rm c}^{(0)} g^{\mu\nu}
= 8 \pi G T^{\mu\nu},
\label{E-eq}
\end{eqnarray}
where $\varLambda_{\rm c}^{(0)}$ is a bare cosmological constant.
From (\ref{T-v}) and (\ref{E-eq}),
the cosmological constant is given by 
$\varLambda_{\rm c} = \varLambda_{\rm c}^{(0)} + \varLambda_{\rm v}$,
effectively.

{\it There can be various sources of $\rho_{\rm v}$.}

The first one is a zero point energy of each particle.
For a relativistic bosonic particle with a mass $m$, 
the energy density $\rho_{\rm z}$ and the pressure $p_{\rm z}$
due to zero point fluctuations are given by
\begin{eqnarray}
\rho_{\rm z} = \frac{1}{2} \int \frac{d^3k}{(2\pi)^3}
\sqrt{\bm{k}^2 + m^2},~~
p_{\rm z} = \frac{1}{6} \int \frac{d^3k}{(2\pi)^3}
\frac{\bm{k}^2}{\sqrt{\bm{k}^2 + m^2}},
\label{zero-point}
\end{eqnarray}
respectively.
Here $\bm{k}$ is a momentum of particle.

The second one is the energy density from the Higgs potential
after the breakdown of electroweak symmetry,
and its absolute value is estimated as
\begin{eqnarray}
|\rho_{\rm Higgs}| = O(v^4) \simeq 10^9 {\rm GeV}^4,
\label{rho-Higgs}
\end{eqnarray}
where $v(=246$GeV) is the vacuum expectation value on the neutral
component of Higgs doublet.

The third one is the energy density accompanied with 
the chiral symmetry breaking due to quark condensations,
and its absolute value is estimated as
\begin{eqnarray}
|\rho_{\rm QCD}| = O(\varLambda_{\rm QCD}^4) 
\simeq 10^{-2} \sim 10^{-3} {\rm GeV}^4,
\label{rho-QCD}
\end{eqnarray}
where $\varLambda_{\rm QCD}$ is the QCD scale.

\subsection{\it Zero point energy density}

{\it The vacuum energy density can receive
large radiative corrections including a cutoff scale.}
The zero point energy density is calculated by using
an effective potential at the one-loop level,
and it naively contains quartic, quadratic and logarithmic
terms concerning an ultra-violet (UV) cutoff parameter $\varLambda$.

By imposing the relativistic invariance of vacuum (\ref{Rel-v})
on $\rho_{\rm z}$ and $p_{\rm z}$, $\rho_{\rm z}$ in (\ref{zero-point}) 
should be of the form~\cite{Akh,K&P},
\begin{eqnarray}
\rho_{\rm z} = \frac{m^4}{64 \pi^2} \ln\frac{m^2}{\varLambda^2},
\label{rho-z}
\end{eqnarray}
up to some finite terms.
Note that the terms proportional to $\varLambda^4$ 
and $m^2 \varLambda^2$
do not satisfy $\rho_{\rm v} = - p_{\rm v}$,
and they can be regarded as artifacts of the regularization procedure.
After the subtraction of logarithmic divergence,
$\rho_{\rm z}$ is given by
\begin{eqnarray}
\rho_{\rm z} = \frac{m^4}{64 \pi^2} \ln\frac{m^2}{\mu^2},
\label{rho-z-mu}
\end{eqnarray}
where $\mu$ is a renormalization point.

For the Higgs boson, its zero point energy is estimated as
\begin{eqnarray}
\rho_{\rm z(Higgs)} \simeq 10^{7} {\rm GeV}^4,
\label{rho-zero-Higgs}
\end{eqnarray}
where we use $m_{\rm h} \doteqdot 126$GeV
for the Higgs boson mass
and take $\mu \simeq 2.4 \times 10^{-13}$GeV
corresponding to the temperature of present universe
$T_0 = 2.73$K.

\subsection{\it Dark energy density}

{\it The experimental value of cosmological constant is estimated
as $\varLambda_{\rm c(exp)} 
= 8 \pi G \cdot 2.4 \times 10^{-47}${\rm GeV}$^4$,
from the observation that the expansion of our present universe 
is accelerating.}

The vacuum energy density of universe is theoretically given by
\begin{eqnarray}
\rho_{\rm v} = \sum_{i} \rho_{{\rm z}(i)}
+ \rho_{\rm Higgs} + \rho_{\rm QCD} + \cdots,
\label{rho-th}
\end{eqnarray}
where $\rho_{{\rm z}(i)}$ is the zero point energy density due to 
a particle labeled by $i$
and the ellipsis stands for other contributions
containing unknown ones from new physics.

From (\ref{rho-Higgs}), (\ref{rho-QCD}) and (\ref{rho-zero-Higgs}),
we have the inequalities,
\begin{eqnarray}
|\rho_{\rm Higgs}| > \rho_{\rm z(Higgs)} \gg |\rho_{\rm QCD}|
\gg \rho_{\rm DE} = 2.4 \times 10^{-47} {\rm GeV}^4,
\label{rho-ineq}
\end{eqnarray}
where $\rho_{\rm DE}$ is a dark energy density 
defined by $\rho_{\rm DE} \equiv \varLambda_{\rm c(exp)}/(8 \pi G)$.
There is a possibility that 
the magnitude of $\rho_{\rm v}$ becomes 
the 4-th power of the terascale
through a cancellation among various contributions
from a higher energy physics based on a powerful symmetry
such as supersymmetry.
Because supersymmetry cannot work
to reduce $\rho_{\rm v}$ close to $\rho_{\rm DE}$,
an unnatural fine-tuning is most commonly
required to realize $\rho_{\rm DE}$.

From (\ref{rho-th}) and (\ref{rho-ineq}),
we have a puzzle that consists of unfitted pieces.
Nature proposes us a big riddle
$\lq$why is the observed vacuum energy density so tiny
compared with the theoretical one?'
and a big mystery 
$\lq$what is an identity of dark energy density?'

\section{\it Candidate of dark energy}

To uncover a clue of CCP and probe into an identity of $\rho_{\rm DE}$,
let us start with the question whether 
$\rho_{\rm v}$ in (\ref{rho-th}) exists in physical reality, it gravitates
or the classical gravitational field feels $\rho_{\rm v}$.

In the absence of gravity,
the vacuum energy from matters $\langle V \rangle$ itself is not observed
directly because there is a freedom to shift the origin of energy.
Here, matters mean various fields 
including radiations such as photon except for graviton. 
For instance, the zero point energy of free fields is removed
by taking a normal ordering in the Hamiltonian.
Only energy differences can be physically meaningful,
as suggested by the Casimir effect.\footnote{
It is also pointed out that the Casimir effect can be formulated
and the Casimir force can be calculated
without reference to the zero point energy~\cite{Jaffe}.}

In the presence of gravity, 
if $\rho_{\rm v}$ gravitates,
the motion of the planets in our solar system
can be affected by $\rho_{\rm v}$~\cite{Wright,CCP2}.
From the non-observation of such an effect for Mercury,
we have a constraint,
\begin{eqnarray}
|\rho_{\rm v}| \le 3 \times 10^{-32} {\rm GeV}^4.
\label{rho-v-const}
\end{eqnarray}
As seen from (\ref{rho-Higgs}), (\ref{rho-QCD}) and (\ref{rho-zero-Higgs}),
the existence of $\rho_{\rm Higgs}$, $\rho_{{\rm z}(i)}$
(for particles heavier than 10eV) 
and/or $\rho_{\rm QCD}$ threatens the stability of our solar system.
Hence,
it seems to be natural to suppose that
the classical gravitational field does not feel
a large portion of $\langle V \rangle$.

In contrast, the ratios of the gravitational mass
to the inertial mass stay for heavy nuclei,
and hence it is reasonable to conclude
that the equivalence principle holds with accuracy
at the atomic level and
the gravitational field couples to 
every process containing radiative corrections,
accompanied by an emission and/or an absorption of matters~\cite{B&P}.
More specifically, the external matter-dependent part of energies must
gravitate in both macroscopic and microscopic world.

Now, let us move to the next step, 
as phenomenological ingredients of CCP are already on the table.

First, based on a standpoint that the Einstein gravity is 
a classical effective theory, 
physics on the CCP can be described by
\begin{eqnarray}
S_{\rm cl} = \int d^4x \sqrt{-g} \left[ 
\frac{1}{16 \pi G} R + \mathscr{L}_{\rm cl}
- \rho_{\rm DE}\right],~~
\rho_{\rm DE} \equiv \varLambda_{\rm c(exp)}/(8 \pi G)
= 2.4 \times 10^{-47} {\rm GeV}^4,
\label{Scl}
\end{eqnarray}
where $R$ is the Ricci scalar made of the classical gravitational field
$g_{\mu\nu}(x)$, $g = \det g_{\mu\nu}$,
$\mathscr{L}_{\rm cl}$ is the Lagrangian density
of matters as classical objects 
and $\mathscr{L}_{\rm cl}$
does not contain a constant term.
Because an effective theory is, in general, an empirical one,
it would not be so strange even if it cannot answer the questions
why a large portion of $\langle V \rangle$ does not gravitate
and what the identity of dark energy is.
Those questions remain as subjects in a fundamental theory.

Second, we explain why it is difficult to derive (\ref{Scl})
in the framework of ordinary quantum field theory,
starting from the action,
\begin{eqnarray}
S = \int d^4x \sqrt{-\hat{g}} \left[ 
\frac{1}{16 \pi G} \hat{R} + \mathscr{L}_{\rm SM}
+ \mathscr{L}_{\alpha}\right],
\label{S}
\end{eqnarray}
where $\hat{R}$ is the Ricci scalar made of the graviton
$\hat{g}_{\mu\nu}(x)$, $\hat{g} = \det \hat{g}_{\mu\nu}$, 
and $\mathscr{L}_{\rm SM}$ and $\mathscr{L}_{\alpha}$
are the Lagrangian densities
for the standard model particles and other particles 
beyond the standard model, respectively.

The amplitudes representing 
the coupling between gravitons and the vacuum energy
are evaluated by calculating Green's functions,
\begin{eqnarray}
G^{(n)}(x_1, \cdots, x_n)
= \langle 0 |{\rm T}(\hat{g}_{\mu_1\nu_1}(x_1) 
\cdots \hat{g}_{\mu_n\nu_n}(x_n))| 0 \rangle.
\label{Gn}
\end{eqnarray}
For example, on the background Minkowski spacetime,
two-point function is written as
\begin{eqnarray}
&~& G^{(2)}(x_1, x_2)
= \frac{i}{4}\int d^4x
\langle 0 |{\rm T}(\hat{h}_{\mu_1\nu_1}(x_1)
{\hat{h}_{\alpha}}^{\beta}(x) )| 0 \rangle
\langle 0 |{\rm T}(\hat{h}_{\mu_2\nu_2}(x_2)
{\hat{h}_{\beta}}^{\alpha}(x) )| 0 \rangle
\langle 0 |{\rm T}(\mathscr{H}'_{\rm int}(x))| 0 \rangle
\nonumber \\
&~& ~~~~~~~~~~~~~~~~~ + \cdots,
\label{G2}
\end{eqnarray}
where $\hat{h}_{\mu\nu}(x)$ is the quantum part of graviton 
in the interaction picture and
$\mathscr{H}'_{\rm int}(x)$ is the interaction Hamiltonian density.
The transition amplitude is obtained by removing
the propagators on the external lines.
The vacuum expectation value
$\langle 0 |{\rm T}(\mathscr{H}'_{\rm int}(x))| 0 \rangle$
corresponds to the vacuum energy density, and
then a large cosmological constant term is derived
after the identification of classical gravitational fields
for external gravitons.
Here, external gravitons mean gravitons in real states
represented by wave functions.

Third, to reconcile (\ref{Scl}) and (\ref{G2}),
we need a radical idea and take a big assumption that
{\it the classical gravitational fields
do not couple to a large portion of the vacuum energy effectively,
in spite of the coupling between graviton and matters
at a microscopic level.}
We expect that it stems from 
unknown features of external gravitons.
For example, if a kind of exclusion principle works,
as a bold hypothesis, that
{\it external gravitons cannot take the same place
in the zero total four-momentum state},
external gravitons would not feel the vacuum energy.
However, if it holds in the strong form,
we would arrive at undesirable conclusions such as
the violation of the equivalence principle
for external matter fields with the zero total four-momentum,
the vanishing scattering amplitudes among only gravitons 
and the absence of dark energy.
To improve them, we need another assumption
such that {\it the exclusive attribute of external gravitons
is violated by the coupling of external matter fields 
(at the same point and/or different ones),
gravitons with derivatives or internal gravitons.}
Here, internal gravitons mean gravitons in virtual states
represented by propagators.
Note that the exclusion principle is merely an example
of reasoning to justify the first assumption.
The point of the second one is that
{\it external gravitons can couple to gravitational
corrections of vacuum energy involving internal gravitons.}

Under the above assumptions,
we give a conjecture on a candidate of dark energy
for the case that the Minkowski spacetime is taken
as a background one,
i.e., $\langle \hat{g}_{\mu\nu}(x) \rangle = \eta_{\mu\nu}$.
In this case, the full propagator of graviton is proportional to
$i/(p^2 - 2 V/M^2 + i \varepsilon)$.
Here, $M$ is the gravitational scale (the reduced Planck scale) defined by 
$M \equiv 1/\sqrt{8 \pi G} (\doteqdot 2.4 \times 10^{18}$GeV).
Using the propagator,
we obtain the gravitational corrections of $V$,
\begin{eqnarray}
\delta_{\rm gr} V = \frac{5}{32 \pi^2} 
\left(\frac{2 V}{{M}^2}\right)^2
\ln \frac{2 V/M^2}{\Lambda^2},
\label{gr-loopV}
\end{eqnarray}
at one-loop level.
By replacing $V$ into $\langle V \rangle$,
we obtain the zero point energy density of $\hat{g}_{\mu\nu}$,
\begin{eqnarray}
\rho_{\rm z(gr)} = \frac{5}{32 \pi^2} 
\left(\frac{2 \langle V \rangle}{{M}^2}\right)^2
\ln \frac{2 \langle V \rangle/{M}^2}{\mu^2},
\label{gr-loop}
\end{eqnarray}
after the subtraction of logarithmic divergence.\footnote{
The study on effective potential 
including contributions of graviton
has been carried out with several motivations,
i.e., the breakdown of symmetry via graviton~\cite{Smolin},
the application to the Vilkovisky-De Witt formalism~\cite{Odintsov,Cho},
the effect of finite tempareture~\cite{B&M}
and the stability of Higgs potential~\cite{L&P,AH&I}.
} 

If $\rho_{\rm z(gr)}$ dominates $\rho_{\rm DE}$,
the magnitudes of $2\langle V \rangle/M^2$ and $\langle V \rangle$
are estimated as
\begin{eqnarray}
\frac{2 \langle V \rangle}{{M}^2} 
\simeq 1.9 \times 10^{-23}{\rm GeV}^2 \simeq
(4.4 \times 10^{-3}{\rm eV})^2
\label{<V>/M2}
\end{eqnarray}
and
\begin{eqnarray}
\langle V \rangle \simeq 5.5 \times 10^{13}{\rm GeV}^4 \simeq
(2.7 {\rm TeV})^4,
\label{<V>}
\end{eqnarray}
respectively.
Here we take $\mu \simeq 2.4 \times 10^{-13}$GeV.
Then, we have a conjecture 
that physics around the terascale is relevant to 
the dark energy of our universe.
If the zero point energy of some scalar field dominates
$\langle V \rangle$, such scalar field has a mass of $O(1)$TeV
and become a candidate of dark matter called 
$\lq\lq$WIMP (Weakly Interacting Massive Particle)''.
If superpartners appear around the terascale,
they can produce zero point energies of $O(1)$TeV$^4$.
In this case, we obtain an interesting scenario that
a vacuum acquires an energy of $O(1)$TeV$^4$
from zero point energies of dark matter and/or superpartners, 
it does not gravitate directly and
the zero point energy of graviton
becomes a source of dark energy. 

\begin{figure}[h!]
\begin{center}
\includegraphics[width=100mm]{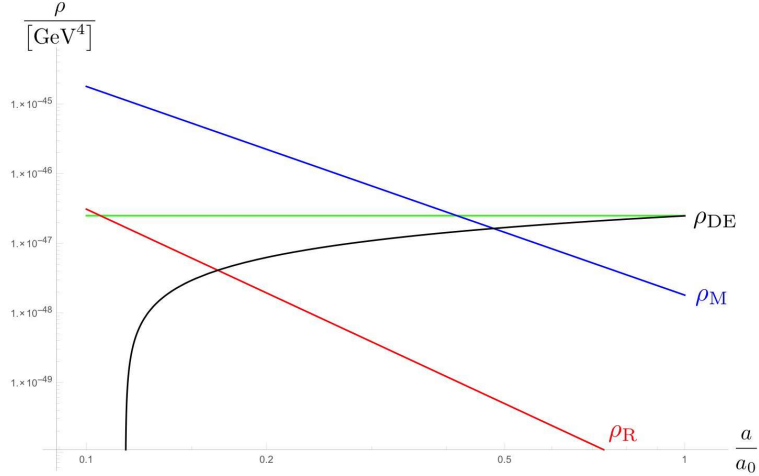}
\vspace{-5mm}
\vskip-\lastskip
\caption{The evolution of several energy densities.
The red, blue and black lines stand for
the evolution of $\rho_{\rm R}$, $\rho_{\rm M}$ and $\rho_{\rm DE}$,
respectively.
Here, $\rho_{\rm R}$ and $\rho_{\rm M}$ are energy densities of
radiations and non-relativistic matters (except for dark matter), respectively.}
\label{F1}
\end{center}
\end{figure}

Let us study the evolution of $\rho_{\rm DE}$
in the case with $\rho_{\rm DE} = \rho_{\rm z(gr)}$
and $\langle V \rangle \simeq (2.7 {\rm TeV})^4$.
If we identify $\mu$ with a temperature of the universe,
$\rho_{\rm DE}$ is not constant but varying 
logarithmically such that
\begin{eqnarray}
\rho_{\rm DE} = \frac{5}{32 \pi^2} 
\left(\frac{2 \langle V \rangle}{{M}^2}\right)^2
\ln \left[\frac{2 \langle V \rangle/{M}^2}{T_0^2}
\left(\frac{a}{a_0}\right)^2\right],
\label{rho-DE}
\end{eqnarray}
where $T_0 = 2.73$K and
$a$ $(a_0)$ is a scale factor of the universe (the present one).
Here, we also use the fact that the temperature is inversely
proportional to the radius of our universe.
The evolution of several energy densities are depicted in Figure \ref{F1}.
\vspace{0mm}
In the appendix, we derive (\ref{rho-DE}) and show
that the logarithmically changing energy density is
described by the equation of state $p = - \rho + {\mbox{constant}}$.

If $\rho_{\rm DE}$ evolves as (\ref{rho-DE}), 
there are several possibilities for physics beyond the terasclae.
First one is that there is no sensitive physics to contribute
the vacuum energy beyond the terascale, i.e.,
no processes associated with a huge vacuum energy
such as the breakdown of grand unified symmetry 
and no superheavy particles generating huge zero point energies.
Second one is that there is a higher energy physics, but
a miraculous cancellation can occur among
various contributions
based on a powerful symmetry such as supersymmetry.
Then, the vacuum energies
can be diminished in supersymmetric grand unified theories
and/or supergravity theories,
and the vacuum energy due to inflaton can
vanish at the end of inflation in early universe.
Third one is that there survives a huge vacuum energy
originated from a higher energy physics
in the absence of a powerful mechanism, but
even virtual gravitons do not
couple to such a vacuum energy beyond the terascale.
Then, another graviton could be required
to realize a higher energy physics.

Finally, we pursue a last possibility
in the presence of the zero point energy of inflaton
with a mass of $O(10^{13})$GeV.
We consider the action,
\begin{eqnarray}
\tilde{S} = \int d^4x \sqrt{-\tilde{g}} \left[ 
\frac{1}{16 \pi \tilde{G}} \tilde{R} + \mathscr{L}_{\rm inf}
+ \cdots\right],
\label{tildeS}
\end{eqnarray}
where $\tilde{G}$ is a coupling constant,
$\tilde{R}$ is the Ricci scalar made of
another graviton $\tilde{g}_{\mu\nu}$,
$\tilde{g} = \det\tilde{g}_{\mu\nu}$,
and $\mathscr{L}_{\rm inf}$ is the Lagrangian density of inflaton.\footnote{
A ghost-free theory could be constructed 
based on the bimetric theory of gravity\cite{H&R}.}
We assume that the ordinary graviton does not couple to
inflaton directly.
In the same way as the ordinary graviton,
we obtain the zero point energy of $\tilde{g}_{\mu\nu}$,
\begin{eqnarray}
\rho_{\rm z(\tilde{gr})} = \frac{5}{32 \pi^2} 
\left(\frac{2 \langle V_{\rm inf} \rangle}{\tilde{M}^2}\right)^2
\ln \frac{2 \langle V_{\rm inf} \rangle/\tilde{M}^2}{\mu^2},
\label{tildegr-loop}
\end{eqnarray}
where $\tilde{M}  = 1/\sqrt{8\pi \tilde{G}}$.
If $\rho_{\rm z(\tilde{gr})}$ dominates $\rho_{\rm DE}$
with $\langle V_{\rm inf} \rangle = O(10^{4 \times 13})$GeV$^4$,
the magnitude of $\tilde{M}$ is estimated as 
$\tilde{M} \simeq 10^{38}$GeV. 

\section{\it Conclusions and discussions}

We have studied physics on the CCP
in the framework of effective field theory 
and suggested that a dominant part of dark energy can originate from
gravitational corrections of vacuum energy
of $O(1)$TeV$^4$,
under the following assumptions. 

\begin{itemize}
\item The graviton $\hat{g}_{\mu\nu}$ couples to 
the potential of matter fields with the strength of $16 \pi {G}$, 
and couples to the vacuum energy of matters
with the same strength in the virtual state.

\item The classical gravitational fields
do not couple to a large portion of the vacuum energy effectively.

\item External gravitons can couple to gravitational
corrections of vacuum energy involving internal gravitons.
\end{itemize}

The second one is beyond our comprehension,
because it is difficult to understand it
in the framework of ordinary local quantum field theory.
However, if the CCP is a highly non-trivial problem 
that cannot be solved without a correct theory of gravity 
in a proper manner, 
it would not be so strange that it
is not explained from the present form of quantum gravity theory.
It ease a major bottleneck of the CCP 
if realized with unknown features of gravity
at a more fundamental level,
and hence we have tried to step boldly from common sense.
As an example of reasoning, we have presented
a kind of exclusion principle that
external gravitons cannot take the same place
in the zero total four-momentum state,
unless they couple to external matter fields (at the same point
and/or different ones),
gravitons with derivatives or internal gravitons.
It may provide a useful hint to disclose physics behind the CCP.

The universe dominated by our dark energy might cause instability,
because it does not satisfy the energy condition 
such as $p + \rho \ge 0$ for perfect fluids.
A phenomenologically viable model must be fulfilled
the requirements
that matters are stable and
a time scale of instability should be longer than the age of universe.

If our speculation were correct,
the CCP is replaced by the challenge
to construct a microscopic theory of gravity
compatible with the above assumptions.
We might need a new ingredient or a novel formalism.
It would be important to pursue much more 
features of graviton and the relationship
between the classical gravitational field and the quantum one.
In such a case, the theory of fat graviton\footnote{
Sundrum has given an interesting proposal
that the cosmological constant is originated from
the vibrational excitations of fat graviton~\cite{Sundrum1,Sundrum2}.
}
may provide a helpful perspective and clue. 

\section*{Acknowledgement}
We would like to thank Prof. T.~Inami,
Prof. K.~Izumi and Prof. P.-M.~Ho
for useful comments on the bimetric theory and the area-metric theory.
According to them, we have returned the title and contents 
of our article to the original one.
\appendix
\section{Evolution of dark energy}

We discuss the evolution of dark energy (\ref{rho-DE}).
The effective potential including contributions of
graviton at one-loop level is given by
\begin{eqnarray}
V_{\rm eff} = V + \frac{5}{32 \pi^2} 
\left(\frac{2 V}{{M}^2}\right)^2
\ln \frac{2 V/{M}^2}{\mu^2},
\label{V-eff}
\end{eqnarray}
where $V$ is the potential of matters.
From the feature that $V_{\rm eff}$ is independent of 
the renormalization point $\mu$,
i.e., $\mu dV_{\rm eff}/d\mu = 0$,
we obtain the relation,
\begin{eqnarray}
\mu \frac{dV}{d\mu} \cdot
\left(1 + \frac{5}{32 \pi^2} \frac{8V}{{M}^4}
\ln \frac{2 V/{M}^2}{\mu^2} + \frac{5}{32 \pi^2} \frac{4V}{{M}^4}
\right) = \frac{5}{32 \pi^2} \frac{8V^2}{{M}^4}.
\label{dV-eff}
\end{eqnarray}
From (\ref{dV-eff}), we find that the magnitude of
$\mu dV/d\mu$ is much less than $\langle V \rangle$
if $\langle V \rangle$ is much less than ${M}^4$.
Then, the zero point energy density
$\rho_{\rm z({gr})}$
given by (\ref{gr-loop}) varies on $\mu$ almost logarithmically.
Hence, the relation (\ref{rho-DE}) is derived
after $\rho_{\rm z({gr})}$ is regarded as 
the dark energy density $\rho_{\rm DE}$
and $\mu$ is identified with the temperature, 
that is proportional to the radius of our universe.

The potential $V$ changes after the incorporation of zero point energies of
particles and the breakdown of symmetry,
and contains $\rho_{\rm v}$ given in (\ref{rho-th}) at present.
From the above observation, the magnitude of $V$
is almost constant except for the period of the early universe.

Next, we show that the logarithmically changing energy density 
is described by the van der Waals type equation of state,
\begin{eqnarray}
p = w \rho - b,~~ w = -1,
\label{eq-state}
\end{eqnarray}
where $b$ is a positive constant.
The energy conservation $d(\rho a^3) = - p d a^3$
is rewritten as
\begin{eqnarray}
d \rho = -(p + \rho) \frac{d a^3}{a^3},
\label{drho}
\end{eqnarray}
where $a$ is the radius of our universe.
From (\ref{eq-state}) and (\ref{drho}), 
we derive the differential equation
$d \rho = b d \ln a^3$ and obtain the solution,
\begin{eqnarray}
\rho = b \ln(a/a_0)^3 + b_0,
\label{rho-b}
\end{eqnarray}
where $b_0$ is a constant.
By comparing (\ref{rho-DE}) with (\ref{rho-b}),
$b$ and $b_0$ are determined as
\begin{eqnarray}
b = \frac{5}{48 \pi^2} 
\left(\frac{2 \langle V \rangle}{{M}^2}\right)^2,~~
b_0 = \frac{5}{32 \pi^2} 
\left(\frac{2 \langle V \rangle}{{M}^2}\right)^2
\ln \left[\frac{2 \langle V \rangle/{M}^2}{T_0^2}\right],
\label{b-b0}
\end{eqnarray}
respectively.

It is pointed out that, for models of dark energy 
characterized by $p = w \rho$,
the system can be unstable if $w < -1$,
which is realized by a negative kinetic term~\cite{CH&T}.
Although our effective theory of dark energy is different from
the case with $w < -1$, $p$ and $\rho$ yielding (\ref{eq-state})
do not satisfy the condition $p + \rho \ge 0$
derived from various energy conditions and might cause instability.
It is not clear whether they are phenomenologically viable or not,
because it depends on the details of model at a microscopic level. 


\end{document}